\documentstyle[aaspp4,12pt]{article}

\newcommand{\be}{\begin{equation}}
\newcommand{\ee}{\end{equation}}
\newcommand{\nn}{\mbox{} \nonumber \\ \mbox{} }
\newcommand{\ba}{\begin{eqnarray}}
\newcommand{\ea}{\end{eqnarray}}

\begin{document}
\title{Inertial Instability of  Accretion Disks with Relativistic Electrons}
\author{Maxim  Lyutikov}
\affil{Canadian Institute for Theoretical Astrophysics,
 St George Street,
Toronto, Ontario, M5S 3H8, Canada}

\begin{abstract}
We show that accretion disks with relativisticly hot electrons which are coupled to cold
ions is subject to  inertial-type instability in which 
 kinematic drag between electrons and ions
produces radial acceleration of electrons.
\end{abstract}

{\it Subject headings}: Accretion Disks - Instabilities

\section{Model Problem}

We consider an accretion disk in which
 electrons are heated to relativistic energies, while
ions remain cold. This condition requires that ratio of electron to ion
heating in the disk 
is larger than $m_e/m_i$ ($m_e$ and $m_i$ are electron and ion masses) 
and that radiation efficiency is relatively small,
 so that a considerable fraction of the gravitational energy released  in accretion
 goes into heat. Ion temperature may still be much larger than electron
temperature.
We also assume that electrons and ions are strongly coupled so that the 
average velocity of the  electron component
is given by the  velocity of ions, which, in turn, is determined by the
mass encompassed by the orbit of an ion at a given radius  $r$. We also require that the
  sound speed $ c_s$ is much less than the
orbital velocity $ c_s \ll \Omega(r) r$, where $\Omega(r)$ is the angular
velocity of the flow. Then 
 we can approximate the flow
as accretion disk with a mean velocity given at each point by the 
velocity of the cold ion fluid in azimuthal direction $ v_{\phi} ^{(0)} = \Omega(r) r$.
The electron component has in addition large thermal spread around the average
velocity:
$  {\bf v} _e = {\bf v} _{\phi} ^{(0)} +  {\bf v} _T$,
where electron thermal velocities $ {\bf v} _T$ are assumed to be
 close to the velocity of light
$  v  _T \sim 1 - {1\over 2 T^2}$ (here $T$ is a relativistic temperature
in units of $m c^2$).
We stress that the assumption that the average electron velocity is  $ v_{\phi} ^{(0)} $
implies that electrons and ions are strongly interacting, i.e. there is a 
drag between electron and ion components. 
This interaction
may not necessarily be through Coulomb collisions but also through collective plasma
modes.

We show that such simple system is unstable toward creation of a large radial current. 
When the current surpasses the critical value of $ j_r \sim e n c_s$
($e$ is a charge of an electron, $n$ is plasma density),
strong  ion sound instabilities develop which would destroy the laminar flow
of ions.

\section{Centrifugal forces}

We 
 concentrate on the  dynamical equations governing electron motion.
 They can be derived from relativistic 
Lagrangian of an electron 
(e.g. Landau-Lifshits, 1951).
\be
{\cal{L}} = - m_e \left( { 1 \over \gamma} + \Phi(r) \right)  , \hskip .3 truein
\gamma = {1\over \sqrt{ 1 - v_r^2 - v_{\theta}^2 - ( v_{\phi} ^{(0)} +  v_{\phi})^2} }
\ee
where $v_r, \, v_{\theta}$  and $ v_{\phi}$ are components of the
random electron velocity, $\Phi(r)= - { G M(r) \over r}$ is a gravitational
potential at a radius $r$ due to the mass $M(r)$ and we set a speed of light to unity.

The equation of motion follows from the Lagrange equation. For the radial component
\be
{d \over d t}  { \partial {\cal{L}}  \over \partial v_r} =
  { \partial {\cal{L}}  \over  \partial r}
\ee
 we find
\ba
&&
\nn
&& 
 { \partial v_r  \over \partial t}  
=    \left( v_{\phi} ^{(0)} +  v_{\phi} \right)
 { \partial  v_{\phi} ^{(0)} \over  \partial r} 
{ \left(1 -  v_r^2 \gamma^2  \right) \over \left(1 +  v_r^2 \gamma^2  \right) }
- {1\over \gamma ( 1 +  v_r^2 \gamma^2 )}  { \partial \Phi(r) \over \partial  r}
\label{dtvr1}
\ea 
where we assumed that $v_{\theta}$  and $v_{\phi}$ are independent of $r$ and $t$.
The first term here is a centrifugal acceleration felt by an electron.

Averaging over random $v_{\phi}$ velocities
in the limit of relativisticly hot electrons, $v_r^2 \gamma^2 \gg 1$ and $v_r^2 \sim 1$, 
 Eq. (\ref{dtvr1}) reduces to
\be
\left\langle  { \partial v_r  \over \partial t}  \right\rangle
= -  {1 \over 2}  { \partial  { v_{\phi} ^{(0)}}^2 \over  \partial r} 
- {1\over \gamma^3}  { \partial \Phi(r) \over \partial  r}
\label{ACC}
\ee
Again,  the first term here is an average centrifugal acceleration felt by electrons.
It is 
 different than the centrifugal force on ions,
$ { { v_{\phi} ^{(0)}}^2 \over  r}$.
This fact may be considered as the main result of this work.

Using the condition that ion motion is a circular motion which satisfies
${ { v_{\phi} ^{(0)}}^2 \over  r} =  { \partial \Phi(r) \over \partial  r}$
Eq. (\ref{ACC}) can be rewritten as
\be
\left\langle  { \partial v_r  \over \partial t}  \right\rangle
= -  v_{\phi} ^{(0)} \left( { v_{\phi} ^{(0)} \over r \gamma^3} + 
{ \partial v_{\phi} ^{(0)} \over \partial r} \right).
\label{ACC1}
\ee
with $\gamma \sim T \gg1$.
Eq. (\ref{ACC1})  shows that  there is a net acceleration
of electrons in radial direction. 

For power law dependence  of $v_{\phi} ^{(0)}$ on radius and for $\gamma \gg 1$ we
can neglect the first term in Eq. (\ref{ACC1}) (which is due to gravitational attraction).
The net radial 
 acceleration of electrons then is purely kinematic:
\be
\left\langle  { \partial v_r  \over \partial t}  \right\rangle
\approx - {1\over 2} { \partial {  v_{\phi} ^{(0)}}^2 \over \partial r}
\label{ACC2}
\ee

 Note, that for solid body rotation, $ v_{\phi} ^{(0)} = \Omega r$ with 
$ \Omega  \sim {\rm const}$ Eq. (\ref{ACC2}) gives  a 
centrifugal force $ { \partial v_r  \over \partial t} = - 
\Omega^2 r$ directed inward (Henriksen \& Norton, 1975, 
Chedia et al. 1996).
Other examples include  Keplerian disk around a central object of mass $ M_{\star}$
for which 
  $ v_{\phi} ^{(0)} =  \sqrt{ { G M_{\star}  \over  r} } $  and 
radial acceleration of  electrons becomes
\be
 { \partial v_r  \over \partial t} =  
 { G M_{\star}  \over 2 r^2}
\ee
Thus, in a Keplerian disk relativistic electrons are accelerated  outwards.
For massive disks the azimuthal velocity 
 $v_{\phi} ^{(0)}$ depends on the mass distribution
in the disk. For  example, if $v_{\phi} ^{(0)}$ is an  increasing function
of  radius
the centrifugal acceleration of   electrons will be directed towards the central object.

\section{Discussion}

 Radial acceleration of electrons with respect to ions will result in a 
formation of a radial current. Current carrying plasmas are susceptible
to various instabilities (e.g. Melrose 1980).
 Similarly to the case of plasma in electric field, 
if the relative motion of electrons and ions is larger than sound velocity
the plasma
 will be  unstable to low frequency
ion sound instabilities with typical growth rates (Melrose 1980)
\be 
\Gamma \sim k c_s  
\ee
 Development of these instabilities will result
in excitation of low frequency  ion sound plasma turbulence. 
Turbulence  will limit, due to the  anomalous viscosity,
 the
relative ambipolar drift between electrons and ions. 
In the extreme case (analogues to Dreicer critical  electric field)
the current in the plasma may reach its maximum value of $ j_{max} \sim e n  c_s $.
This is a large global current that flows in radial direction.

There are several implications of the instability.
First, there is a generation of the low frequency 
turbulence that would destroy the laminar flow of ions and may contribute to the
angular momentum transport in the disk. 
Secondly, the radial current
 will  produce azimuthal magnetic
field of the order $B_{\phi} \sim j_{max} R$ which may play an important role
both for the  magneto-rotational instabilities  (Balbus \& Hawley 1991) 
and overall magnetospheric structure
of the system. Thirdly, the
radial current will try to charge the central
object.  
The implications of this current on  the global structure of the disk 
are not clear.

Finally, we give a "physical reason" for the unusual
expression for the centrifugal acceleration of  electrons.
As the electron is displaced outward in radial direction the
loss in the potential energy in the effective potential
(Landau \& Lifshitz 1976)
is accompanied  by the change of  the kinetic energy of azimuthal motion.
For $ \Omega(r)$  falling off slower than $r^{-1}$ the gain in the 
kinetic energy is larger than  the loss of the potential energy, so that
electron "prefers" to go inward - centrifugal force reverses.

\acknowledgements
 I  would like to thank Norm Murray
 for useful comments.

\end{document}